\documentstyle[aps,epsfig,twocolumn]{revtex}

\global\firstfigfalse

\begin{document}
\twocolumn[
\hsize\textwidth\columnwidth\hsize\csname @twocolumnfalse\endcsname

\title{$^{139}$La Spectrum and Spin-Lattice Relaxation Measurements of 
La$_{2/3}$Ca$_{1/3}$MnO$_3$ in the Paramagnetic State}

\author{K.E. Sakaie, C.P. Slichter, P.Lin, M. Jaime\cite{JAIMELANL}, M.B. Salamon}

\address{Physics Department and Frederick Seitz Materials Research Laboratory\\
University of Illinois at Urbana-Champaign\\ 1110 W. Green Street, Urbana, IL 61801-3080}

\maketitle
\begin{abstract}
This paper reports $^{139}$La NMR measurements of a powder sample of the colossal 
magnetoresistance compound La$_{2/3}$Ca$_{1/3}$MnO$_3$ (T$_c$ =268K) performed in the paramagnetic state (292K-575K) and in high magnetic fields (2.00-9.40 Tesla).  Analysis of the spectrum measured at 575K 
establishes that the spectrum is a standard powder pattern broadened to a significant degree by a variation in lattice distortions around lanthanum nuclear sites. At lower temperatures, but still above T$_c$, the spectrum shifts and broadens.  Both the shift and broadening exhibit Curie-Weiss behavior, indicating that the shift measures the polarization of the electron spin system, and the broadening reflects a distribution of magnetic susceptibilities.  This distribution may result from variations of local susceptibility in the bulk of the sample or from differences in demagnetizing factors among powder grains.  Close inspection of the spectrum indicates that the lattice distortions do not change as the temperature lowers.  Spectral diffusion measurements suggest that the temperature dependence of the spectrum shape does not result from freezing out of motion of magnetic polarons.  Variations in the nuclear spin-lattice relaxation across the spectrum indicate that magnetic fluctuations, not lattice vibrations, dominate nuclear relaxation.  Nuclear spin-lattice relaxation therefore measures electron spin dynamics in this system.  The magnetic field dependence of the spin-lattice relaxation indicates that the electron spin-spin correlation function adopts simple single exponential behavior with a slow field-independent correlation time of 10$^{-8}$ seconds near T$_c$.  The spin-spin correlation function changes form at higher temperatures and can be described by introducing a field dependence to the correlation time and to the magnitude of the fluctuating field.  Even at the highest temperatures, the correlation time remains slow, on the order of 10$^{-9}$ seconds.  The spin-lattice relaxation therefore indicates the presence of extremely slow dynamics above T$_c$.
\end{abstract}
\narrowtext]
\section{Introduction}
Colossal magnetoresistance manganites provide an opportunity to study systems in which strong 
coupling among charge, spin, orbital, and lattice degrees of freedom determine bulk properties \cite{MILLIS}.  Many experiments indicate the presence of local entities in these compounds. Neutron scattering experiments find small magnetic clusters in the paramagnetic state for which the correlation length does not diverge at T$_c$ \cite{DETERESA}.   A large diffusive peak seen by inelastic neutron 
scattering \cite{LYNN,BACA}, and a broad distribution of muon spin relaxation rates \cite{HEFFNER} indicate that clusters persist far below the Curie temperature.  The dynamics and trapping of these clusters may be responsible for the large magnetoresistance and other unusual transport and magnetic properties.  Neutron pair distribution function \cite{BILLINGE,LOUCA} and X-ray absorption fine structure measurements \cite{BOOTH,TYSON} demonstrate the presence of local distortions of MnO$_6$ octahedra in which the Mn-O bond lengths differ significantly from the average value.  While these distortions change during the nearly simultaneous ferromagnetic and metal-insulator transitions, they persist into the metallic regime.  The precise role these clusters and distortions play in the behavior of the manganites remains unclear.

Because characterization of local properties is needed to understand the manganites, information 
from a local probe such as Nuclear Magnetic Resonance (NMR) should be valuable.  NMR measurements 
determine properties at a well-defined point in the system-- the nuclear site-- but do not require that the system exhibit long range order.  In fact, NMR can often characterize the degree of disorder in a system. Analysis of the NMR spectrum determines the size and distribution of static local magnetic susceptibilities and lattice distortions.  NMR spin-lattice relaxation rate measurements characterize extremely low energy fluctuations (on the order of 0.1$\mu$eV) of electron spins or of the lattice.

This paper presents $^{139}$La NMR spectrum and nuclear spin-lattice relaxation measurements of\\ La$_{2/3}$Ca$_{1/3}$MnO$_3$ (T$_c$=268K).  Measurements were performed in the paramagnetic state (292K-575K) in high magnetic fields (2.00-9.40 Tesla).  The sample is a powder synthesized by a standard solid state procedure.  Chemical analysis found no traces of contaminants.  T$_c$ was determined by magnetization measurements.  The magnetization transition is sharp, the full width at half maximum of the $\frac{dM}{dT}$ curve at T$_c$ in a 200 Oe field being 8K.  The width of the transition compares favorably with that measured in other ceramic samples \cite{HWANG} and indicates that the sample is single phase with small deviation from the main composition.

The paper begins with an introduction to general properties of a $^{139}$La NMR spectrum.  An 
analysis of the high temperature data demonstrates that the spectrum is a standard powder pattern, 
broadened to a significant degree by a variation in lattice distortions around different lanthanum sites.  As the temperature lowers towards T$_c$, the spectrum shifts and broadens due to the onset of ferromagnetism.  The shift is proportional to the electron spin polarization while the broadening reflects a distribution of susceptibilities.  Close inspection of the spectrum indicates that the lattice distortions do not change in the temperature range studied.  Spectral diffusion measurements show no signs of the freezing out of motion of magnetic polarons.  After an introduction to some of the principles of nuclear spin relaxation, we provide evidence that magnetic fluctuations, and hence electron spin dynamics, cause the relaxation.  Finally, the field dependence of the nuclear spin-lattice relaxation reveals unusual and slow electron spin dynamics in the paramagnetic state.

\section{MEASUREMENTS AND DISCUSSION}	
The $^{139}$La nucleus has a spin I=$7/2$ and a magnetic dipole moment $\mu=\gamma_n\hbar\hat{I}$, where $\gamma_n$ is the gyromagnetic ratio of the nucleus.  For $^{139}$La, $^{139}\gamma/2\pi$=6.014MHz/Tesla. The interaction between the magnetic dipole moment and a magnetic field, B, creates 2I+1 energy levels with energies $E_m=m\gamma_n\hbar B$, where m may adopt the values $I,I-1,I-2,\ldots,-I$.  NMR measurements detect transitions between energy levels m, m' such that $|m-m'|$=1 with frequencies, $\omega_0$, proportional to the magnetic field at the nucleus:
\begin{equation}
\omega_0 = \gamma_n B.
\end{equation}
Because the gyromagnetic ratio is known to a high degree of accuracy, the resonant frequency measures 
the magnetic field at the nuclear site. 

In a paramagnet, the field, B, at the nucleus is given by $B=(1+K)H_0$, where K, the fractional frequency shift of the spectrum, is proportional to the magnetic susceptibility at the nuclear site, and $H_0$ is the applied magnetic field.  Measurement of the shift of the spectrum thus yields, within a constant of proportionality, the local magnetic susceptibility.  The precise  relation between the shift and the susceptibility depends on the properties of the system.  The shift may result from intrinsic properties, such as the electron hyperfine field, and from bulk effects, such as demagnetizing fields.  A distribution of shifts broadens the spectrum by an amount proportional to the product of the width of the distribution and the magnetization.  The distribution may result from a variation in electronic structure among sites, variations in demagnetizing fields among differently shaped grains in a powdered sample, or, in a randomly oriented powder, from anisotropy of the shift.  

The $^{139}$La nucleus also possesses an electric quadrupole moment, Q=$20\times10^{-30}$meter$^2$, which 
interacts with electric field gradients.  Electric field gradients result from the distribution of charges around the nucleus.  The electric quadrupole interaction creates structure in the NMR spectrum.  A site with cubic symmetry has an electric field gradient of zero, so the size of the effect of the electric quadrupole interaction on the NMR spectrum measures the degree of deviation from cubic symmetry of the surroundings of the nuclear site. 

For the measurements described here, the electric quadrupole interaction may be considered a perturbation of the magnetic interaction.  Under the influence of  the magnetic and the electric quadrupole interaction, the NMR spectrum exhibits several sharp peaks in an ordered single crystal.  The frequency of the peak corresponding to the transition between level m and m-1 is:
\begin{equation}
\omega_m=\gamma_n(1+K)H_0 + \frac{3}{2}\frac{e^2qQ}{4\hbar I(2I-1)}(3\cos^2\theta-1)(2m-1).
\end{equation}
The first term results from the magnetic interaction.  The second is a shift due to the electric quadrupole interaction taken to first order in perturbation theory.  $e$ is the charge of the electron.  $q$ is the strength of the electric field gradient.  $\theta$ is the angle between the principal axis of the electric field gradient and the magnetic field.  This formula applies if the electric field gradient is axially symmetric.  

According to the above equation, the transition between the $m=1/2$ and $m=-1/2$ levels is 
unaffected by the electric quadrupole interaction to first order.  This transition is, however, shifted to second order by an amount given by:
\begin{equation}
\Delta\omega^{(2)}_{1/2}=\frac{9}{64}\frac{4I^2}{2I-1}\left(\frac{e^2qQ}{\hbar}\right)^2\frac{(1-9\cos^2\theta)(1-\cos^2\theta)}{\gamma_n(1+K)H_0}
\end{equation}
The electric quadrupole interaction affects the transition between the $m=1/2$ and $m=-1/2$ levels only 
slightly but shifts the other transitions by larger amounts to higher and lower frequencies.  The $+1/2$ to $-1/2$ transition is known as the ``central transition" while the other transitions are known as the ``satellites." The magnetic field dependence of the quadrupole interaction is particularly important.  Analysis of the spectrum in this paper utilizes the fact that the first order shift is independent of magnetic field whereas the second order shift varies inversely with magnetic field.
	
The crystal axes determine the orientation of the electric field gradient.  In a powder sample, the 
crystal axes, and hence the electric field gradient, adopt a distribution of angles with respect to the 
applied magnetic field.  This angular distribution broadens the sharp peaks of a single crystal into a 
characteristic powder pattern which has been described and calculated in the literature \cite{GERSTEIN}.  
In the powder pattern, the satellites spread out considerably and overlap.  The first order quadrupole 
interaction determines the satellites' width, which is therefore proportional to the size of the electric field gradient and independent of magnetic field.  The average of the first order interaction over all angles is zero, so the average shift of the satellites due to the first order quadrupole interaction is zero.  The satellites do, however, exhibit a shift due to the magnetic interaction.

In contrast,  the central transition, feeling only a second order interaction, is much sharper than 
the satellites.  The second order interaction varies as $q^2/H_0$, so the powder-average width of the central transition is inversely proportional to the magnetic field.  The average of the second order interaction over all angles is nonzero, so the average position of the central transition is shifted, also by an amount inversely proportional to the magnetic field.  The average frequency of the central transition for a spin-$7/2$ is:
\begin{equation}
\omega_{1/2}=\gamma_n(1+K)H_0 - \frac{1}{392}\left(\frac{e^2qQ}{\hbar}\right)^2\frac{1}{\gamma_n(1+K)H_0}.
\end{equation}
The frequency of the central transition thus contains a component proportional to and a component 
inversely proportional to the magnetic field.  The effects of a deviation from axial symmetry (a non-zero asymmetry parameter) on the average frequency are small, being no more than two percent of that due to the axially symmetric part of the quadrupole interaction.

\subsection{The Spectrum at High Temperature}
Figure~\ref{HTSPECTRUM} depicts the $^{139}$La spectra in La$_{2/3}$Ca$_{1/3}$MnO$_3$ at 575K measured at 8.19 and 4.00 Tesla.  The frequency shift is measured with respect to the unshifted resonant frequency derived using the value of $H_0$ determined from the resonant frequency of deuterium in deuterated water.  The spectrum consists of a narrow feature (300 kHz FWHM at 8.19 Tesla, 680kHz at 4.00 Tesla) and a broad feature (4 MHz FWHM at both fields).  The inset of figure~\ref{HTSPECTRUM} compares the 
narrow feature measured at 4.00 Tesla and 8.19 Tesla after multiplying the frequency shift by the value of the field.  After this scaling of the frequency axis, the narrow features lie on top of each other, 
demonstrating that the width scales inversely with field.  We can therefore conclude that the narrow 
feature corresponds to the central transition of  the spectrum of a spin-$7/2$ nucleus.  Resolution of the broad feature at the lower field is limited by a small signal to noise ratio.  However, the broad feature does not become wider upon lowering the field, indicating that it corresponds to the satellites.  Further evidence for this assignment of the broad feature comes from a fit of the spectrum.  

The average shift of the central transition has a component which varies as $q^2/H_0$.  Therefore, 
measurements of the frequency of the center of the narrow feature of the spectrum as a function of applied magnetic field, $H_0$, determine the average strength of the electric field gradient, $q$.  Measurements at fields ranging from 2.00 to 8.19 Tesla determine a value of electric quadrupole coupling, $\nu_{Q}\equiv\frac{e^2qQ}{h}$=26 MHz.  For the sake of comparison, for $^{139}$La in the layered cuprate system LaCuO$_{4+\delta}$, $\nu_{Q}$ = 89 MHz \cite{HAMMEL}.  Unlike the cuprate,  the crystal structure of the manganite is nearly cubic, with a slight orthorhombic distortion \cite{DAI}.  However, the nonzero electric 
field gradient demonstrates the sensitivity of NMR to the deviation from cubic symmetry, which can be 
safely ignored in a number of other types of measurements.

Given the measurement of  the average strength of the electric field gradient, we can fit the 
spectrum.  Figures~\ref{UNBROADFIT}, \ref{MAGBROADFIT}, and \ref{EFGBROADFIT} show several fits, each assuming an electric field gradient with axial symmetry, with insets showing expanded views of the narrow feature and of a portion of the wide feature.  Figure~\ref{UNBROADFIT} depicts a calculated powder pattern spectrum which uses no adjustable parameters.  The amplitude has 
been chosen so that the total area under the calculated powder pattern is the same as the area under the 
measured spectrum.  The narrow feature in the calculated powder pattern is the central transition, and the broader features are the satellites.  The widths of the features of the calculated powder pattern and of the measured spectrum match closely.  We can also compare the areas under the different parts of the spectrum.  In the calculated powder pattern, the ratio of the area under the satellites to the area under the central transition is 4.96.  For comparison, we fit the measured spectrum to a superposition of a broad and a narrow gaussian.  The ratio of the areas of the gaussians is $5.7\pm0.5$.  Given the difficulties involved with precisely measuring absolute intensities over the large frequency range involved in this experiment, the area ratios show reasonable agreement.

The theoretical spectrum shows sharp peaks which are not found experimentally.  We can 
improve the fit by introducing broadening mechanisms.  Figure~\ref{MAGBROADFIT} shows the effect of broadening by a 
distribution of magnetic shifts.  The central transition of the calculated powder pattern matches the 
narrow feature of the measured spectrum well, but the smeared singularities of the calculated satellites (figure~\ref{MAGBROADFIT}, inset on right) do not appear in the measured spectrum.  Increasing the broadening to make the satellites match the broad feature makes the central transition too wide.  A distribution of magnetic shifts does not by itself account for the spectrum shape.

By including a distribution of electric field gradient as well as magnetic broadening, we can fit 
both the broad and narrow features.  Figure~\ref{EFGBROADFIT} shows a powder pattern with a gaussian distribution of the electric quadrupole coupling of width $\Delta\nu_{Q}/\nu_{Q}$=15\%.  This substantial 
distribution indicates that distortions in the surroundings of different lanthanum sites vary significantly.  The fit also involves a convolution with a gaussian with a width 0.23\% of the average NMR frequency which represents a slight distribution of magnetic shifts.  A different distribution or a deviation of the electric field gradient from axial symmetry would improve the fit.  However, the quality of this simple fit confirms the interpretation of the spectrum as a central transition and satellites. 

The presence of a distribution in lattice distortions agrees with results of other local 
measurements of the lattice \cite{BILLINGE,LOUCA,BOOTH,TYSON} which find a distribution of manganese-oxygen and lanthanum-oxygen distances.  To affect the electric field gradient at the lanthanum site, the distortions need not be in bonds involving the lanthanum nucleus.  Variations in distant manganese-oxygen bonds can have a measurable effect.  Such distortions may result from defects, Jahn-Teller distortions of MnO$_6$ octahedra, or from distortions due to the difference between the radii of La$^{3+}$ and Ca$^{2+}$.  We do not present a quantitative comparison of distortions seen by NMR and those measured by other methods because accurate calculations of electric field gradients require characterization of the charge distribution around the nucleus to a level of precision beyond the scope of this investigation.

One may also reasonably expect that local lattice distortions would cause deviations from axial symmetry.  However, including a distribution of such deviations in the analysis has a negligible effect on the values of the size and distribution of electric quadrupole coupling.  This insensitivity also precludes quantitative characterization of the asymmetry parameter.

Our analysis of the spectrum rules out an interpretation of the structure of the spectrum which 
invokes a number of sharp features corresponding to distinct sites.  In such an interpretation, the narrow feature would, for example, correspond to a single site of high symmetry while the broad feature would correspond to a site of lower symmetry or to a number of 
narrower, overlapping components.  Such an interpretation explains the structure of the spectrum in the 
ferromagnetic state \cite{ALLODI1,ALLODI2}, but does not apply in the paramagnetic state.
\subsection{Temperature Dependence of the Spectrum}

Figure~\ref{TEMPDEP} depicts the results at several temperatures ranging from 292K to 575K.  The spectrum experiences a change in shape and an enormous frequency shift on the order of 20\% at room temperature.  The shift is comparable to the zero field resonant frequency of 20MHz observed in the ferromagnetic state \cite{ALLODI1,ALLODI2}.  In contrast, the bulk susceptibility creates a macroscopic demagnetizing field on the order of 1\% which alone would lead to a small negative shift.  The large shift and sizable zero field frequency arise from the electron-nucleus hyperfine coupling,
\begin{equation}
{\cal H}_{hf}=\vec{I}\cdot\tensor{A}\cdot\vec{S},
\end{equation}
where $\vec{I}$ and $\vec{S}$ are the nuclear and electron spin operators and $\tensor{A}$ is the hyperfine coupling constant.  
Assuming an isotropic hyperfine coupling, the shift adopts the form:
\begin{equation}
K=\frac{A\chi_e}{\gamma_n\gamma_e\hbar^2},
\end{equation}								
where $\chi_e$ is the static electron susceptibility, and $\gamma_n$, $\gamma_e$ are the gyromagnetic ratios of the nucleus and of the electron.  The hyperfine coupling constant, A, provides a measurement of the electron wave function at the nuclear site.  Given the hyperfine coupling constant, the shift gives an independent measurement of the susceptibility.

A plot of the inverse of the shift of the spectrum center of mass versus temperature combined with its magnitude and sign confirms that hyperfine coupling to the electron system determines the shift.  Figure~\ref{CURIE} shows that the shift  obeys Curie-Weiss behavior with a Curie temperature of 269$\pm$19K, in agreement 
with the value of T$_c$=268K determined from magnetization measurements.  The large error bars on the 
Curie temperature result from the small number of data points and also reflect the deviation of this system from simple Curie-Weiss behavior \cite{DETERESA}, a hallmark of short range ferromagnetic correlations.  Nevertheless, the shift clearly depends on the polarization of the electron system.  

Along with the shift, the spectrum changes shape.  As figure~\ref{TEMPDEP} shows, the narrow feature 
disappears as the temperature drops.  However,  superimposing the room temperature spectrum and the 
broad part of the high temperature spectrum (figure~\ref{EFGINDEPT}) indicates that the overall width remains 
unchanged.  As explained earlier, the size of the electric field gradient, a result of static lattice distortions, determines the overall width of the high temperature spectrum.  The similarity between the high temperature and room temperature spectra indicates that, from the viewpoint of the lanthanum site, the size and distribution of lattice distortions change little with temperature in the paramagnetic state. 

As we explain below, the loss of the narrow central feature results from a distribution of susceptibility resulting either from a variation of local electronic structure among nuclear sites or from a distribution of demagnetizing fields among differently shaped powder grains.  The resulting broadening of the narrow feature is proportional to the product of the magnetization and the width of the distribution.  To characterize the size of the broadening, we start from the fit of the spectrum at high temperatures depicted in figure~\ref{EFGBROADFIT}.  The fit includes a slight magnetic broadening calculated by convolution with a gaussian.  We can fit the spectrum at all temperatures by varying the width of the gaussian.  A magnetic distribution sufficient to broaden the narrow feature to match the width of the wide feature affects the wide feature only slightly.  The width of the convolution gaussian exhibits Curie-Weiss behavior (figure~\ref{CURIEWIDTH}) with a Curie temperature of 280$\pm$30K, in agreement with the value found from the temperature behavior of the shift.  Therefore, the broadening is proportional to the magnetization, consistent with a picture in which a distribution in susceptibility drives the temperature dependence of the width of the central transition.

Another possible explanation for the change in the spectrum shape involves changes in the 
hopping rate of small polarons or other local field inhomogeneities.  A number of experiments detect the 
presence of small magnetic polarons in the manganites \cite{DETERESA,JAIMEAPL,JAIMEPRL}.  Associated with the small magnetic polarons are inhomogeneous magnetic fields which could affect the spectrum shape.  Nuclei near the localized moment of the small polaron would experience a larger shift than those far away.  If the polaron is fixed in space, its inhomogeneous field would broaden the spectrum.  If the moment hops at a rate exceeding the root mean squared frequency shift induced by its inhomogeneous field, the nuclear system would exhibit a single, average shift instead of a distribution of shifts.  As a result, the spectrum would narrow with increasing hopping rate.  This phenomenon is known as motional narrowing \cite{SLICHTER}. 

To see if changes in the hopping rates of local moments affect the spectrum, we look for spectral 
diffusion by performing a type of hole burning experiment called an S-Wave measurement, which is described in more detail elsewhere \cite{BECERRA}.   These measurements were performed at the peak of the central transition in a 4 Tesla field at room temperature, 350K, and 575K.  To perform the measurement, we use pulsed NMR techniques to selectively invert one section of the spectrum [figure~\ref{SWAVE}] and watch the spectrum as it evolves over a time, T$_{ev}$, after the inversion.  For the sake of explanation, assume that the signal at the higher frequency corresponds to nuclei near a local moment and the signal at the lower frequency corresponds to nuclei far away.  We start the experiment by inverting the signal, and hence the nuclear magnetization, on the high frequency side of the spectrum.  If the local moment hops while the spectrum recovers, the nuclei originally near the moment will then be far from the moment and hence their signal will move to the lower frequency.  As the inverted magnetization also moves to the lower frequency, a dip in the spectrum would simultaneously appear at the lower frequency.  The situation involving a distribution of local fields is more complicated, but a dip in the signal as time progresses generally indicates the presence of hopping.  In the absence of hopping, the entire spectrum only grows with time via spin-lattice relaxation. Figure~\ref{SWAVE} shows that the magnetization only grows over time, suggesting that hopping does not affect the spectrum at room temperature.  Similar measurements suggest that hopping does not affect the spectrum at 350K or 575K.  

The lack of spectral diffusion in the S-Wave measurement may, however, stem from limitations in the experiment.  First, the excitation bandwidth is at most one-fourth the width of the central transition, so effects of hopping on parts of the spectrum outside the excitation bandwidth would not be detected.  Also, if there is no correspondence between spectrum position and the spatial separation between a nucleus and a polaron, hopping of a polaron in space will not lead to spectral diffusion.  Our measurements thus do not conclusively rule out hopping as a cause for the narrowing of the spectrum with increasing temperature.

Furthermore, even if we accept that small polarons do not contribute to the change in the spectrum shape, such a conclusion would not contradict other measurements which detect the presence of small polarons. The density of charge carriers, and hence of small polarons, is rather large compared to that found in conventional small polaron systems \cite{MOTT}.  Furthermore, DeTeresa {\em et. al.} \cite{DETERESA} find that an applied magnetic field comparable to those used in these experiments increases the correlation length of magnetic clusters.  Due to their density and size, small polarons may not present an inhomogeneous field to the lanthanum sites.  Also, the motion of small polarons may be too fast, over the entire temperature regime studied, to affect the spectrum shape.  The hopping rate in the temperature range studied is, at the slowest, on the order of $10^{11}$Hz \cite{JAIMEPRL}.  Motional narrowing of a spectrum occurs when the hopping rate reaches and exceeds the width of the spectrum which, in this case, is on the order of 4 MHz.  We should therefore expect that changes in the dynamics of small polarons will have no effect on the spectrum shape.  However, the 4 MHz lower bound on the rate of motion of a field inhomogeneity applies to any type of field inhomogeneity--- an assertion independent of models which invoke the presence of small polarons.

\subsection{Relaxation Times}
The NMR spectral shape yields information about static, or nearly static, structural properties of the 
system.  Measurements of nuclear spin-lattice relaxation determine the intensity and correlation times of magnetic and lattice fluctuations.  In typical magnetic systems, magnetic fluctuations dominate the 
relaxation.  The manganites, however, exhibit strong dynamic lattice distortions \cite{DAI} which induce 
fluctuations in the electric field gradient that may dominate the relaxation.  As explained below, 
measurements of nuclear spin-lattice relaxation performed at different points along the spectrum indicate that magnetic fluctuations dominate the relaxation.

We start with a discussion of general aspects of nuclear spin-lattice relaxation of a spin-$7/2$
nucleus.  At equilibrium, the population of nuclear energy levels obeys a Boltzmann distribution.  After 
the populations are disturbed from equilibrium, they approach their equilibrium values in a 
characteristic time known as the spin-lattice relaxation time.  The size of the NMR signal 
depends on the population difference between levels, so the rate at which the NMR signal approaches its 
equilibrium value after a perturbation depends on the spin-lattice relaxation time.  One type of 
measurement of spin-lattice relaxation time is the so-called inversion-recovery experiment, in which the 
populations of two levels are inverted.  The signal, M(t),  associated with these levels is measured at 
several times t after the inversion.  The rate at which M(t) approaches its equilibrium value depends on 
the spin lattice relaxation time.

The intensity of the transition between levels m and m+1 of a spin-7/2 nucleus exhibits multiple-
exponential spin lattice relaxation \cite{ANDREW,MARTINDALE}:
\begin{equation}
M^m(t)\propto\sum_n1-a^m_n\exp\left(-\frac{b^m_nt}{\tau}\right).
\end{equation}					
The spin-lattice relaxation time, $\tau$, depends on the intensity of fluctuations in the magnetic field and electric field gradient at the nucleus.  The constants a$^m_n$, b$^m_n$, depend on the initial conditions of the populations of the different energy levels.  Depending on whether fluctuations of the magnetic field or of the electric field gradient dominate the relaxation, these constants vary with m differently.  Calculations described in the appendix indicate that, for a spin-$7/2$, the multiple-exponential spin lattice relaxation differs only slightly from a single exponential, 
\begin{equation}
M^m(t)\propto1-2\exp\left(-\frac{t}{T^m_1}\right).
\end{equation}
Furthermore, as explained in the appendix, if magnetic fluctuations dominate relaxation, T$^m_1$ is smaller when measured on the central transition than on the satellites.  If electric field gradient fluctuations dominate, the opposite is true.

Measurements of spin-lattice relaxation at different points along a spectrum therefore determine 
the relaxation mechanism. Figure~\ref{T1VSPECT} depicts effective single-exponential relaxation times measured at 
different positions on the spectrum at 9.4 Tesla and at approximately 575K.  The repetition time of this and all subsequent measurements has been set sufficiently long to ensure that the spin system starts at equilibrium before each measurement. The spectrum at 9.4 Tesla is similar to that shown in Figure~\ref{HTSPECTRUM}, except that the width of central feature (265 kHz FWHM) is smaller at higher field.  The values of relaxation times have been normalized to the value measured at the apex of the central transition, and the frequency shift refers to the frequency of the apex.  The spin-lattice relaxation time is shortest at the apex of the central transition, indicating that magnetic fluctuations cause the relaxation. 

Quantitative analysis of the variation of the relaxation at different points along the spectrum 
depends on details of the spectrum shape.  Our analysis of the spectrum indicates that the various 
transitions overlap.  Measurements of spin-lattice relaxation at any particular frequency will therefore 
contain contributions from several transitions at once.  From the parameters used in the fit to the spectrum shown in figure~\ref{EFGBROADFIT}, we can determine the degree of overlap among transitions and so can calculate how the spin-lattice relaxation time varies among the different transitions.  There is an average spin-lattice relaxation time due to contributions from overlapping transitions.  The line in figure~\ref{T1VSPECT} represents the calculated average spin-lattice relaxation time as a function of position along the spectrum.  The agreement between calculation and measurement supports our interpretation of the origin of spin-lattice relaxation.  Since this calculation relies on the parameters used in the fit of the spectrum shape, the agreement also lends further support to our analysis of the spectrum shape.
	
Nuclear spin-lattice relaxation measurements probe magnetic fluctuations and hence 
electron spin dynamics.  The spin-lattice relaxation time is related to the electron spin-spin correlation function via:
\begin{equation}
\frac{1}{\tau}\propto\int dt \cos\omega_0t<h(t)h(0)>
\end{equation}
where $\omega_0$ is the NMR frequency and h(t) is the fluctuating field at the nucleus due to the electron spin.    
The constant of proportionality depends on the magnetic moment of the nucleus and on the transition excited in the experiment. One consequence of the above formula is that the nuclear spin-lattice relaxation time depends on 
fluctuations at the NMR frequency, which are of extraordinarily low energy, on the order of 0.1$\mu$eV in this system.  For the sake of comparison, neutron scattering experiments can resolve fluctuations down to the meV range.  NMR as a probe of spin dynamics thus complements  inelastic neutron scattering.  

If we assume a simple exponential form for the electron spin-spin correlation function:
\begin{equation}
<h(t)h(0)> = h_0^2\exp\left(-\frac{t}{\tau_c}\right),
\end{equation}						
in which $h_0$ is the magnitude of the fluctuating field and $\tau_c$ is the correlation time of the fluctuation, the spin-lattice relaxation time adopts the form \cite{BPP}:
\begin{equation}
\frac{1}{\tau}=\frac{\gamma^2_n h^2_0}{3}\frac{h^2_0\tau_c}{1+\omega^2_0\tau_c^2},
\end{equation}
where the value of the constant applies for measurements at the central transition for a spin-7/2 nucleus. Assuming that $h_0$ and $\tau_c$ are field-independent, the spin-lattice relaxation time varies linearly with the square of the applied magnetic field, $H_0^2$.  Figure~\ref{T1VH} shows that, at room temperature, the field dependence of the spin-lattice relaxation time exhibits such behavior.  Furthermore, we find $h_0=130$ Gauss and $\tau_c=10^{-8}$ second.  These values are only slightly affected by the presence of overlapping transitions in the powder pattern spectrum.  The size of the fluctuating field is the same order of magnitude as the dipolar field of an electron sitting on a near neighbor manganese site, but is several orders of magnitude smaller than the hyperfine field which describes the shift.  The correlation time is also several orders of magnitude slower than that associated with small polaron hopping.  The slow dynamics may be related to that seen in the ferromagnetic state by muon spin relaxation measurements \cite{HEFFNER} which imply spin glass-like behavior of magnetic clusters.  Electron spin resonance measurements, however, do not see anomalous line broadening above T$_c$ which one would associate with glassy dynamics \cite{LOFLAND}. 
	
At higher temperatures, the field dependence of the nuclear spin-lattice relaxation time changes.  
At 375K, the spin-lattice relaxation time is proportional to $H_0$, not $H_0^2$.  The simplest possible 
explanation for such behavior involves the same correlation function used above but with field dependent 
parameters:  $\tau_c\propto H^{-1/2}_0$ and $h_0\propto H^{1/4}_0$.   Due to limits on the signal to noise ratio, we cannot determine whether the spin-lattice relaxation time at highest measured temperature, 575K, varies as $H_0$ or H$_0^2$.  However, regardless of the form used to fit the data, the correlation time at the higher temperatures is faster than that at room temperature, but still slow, on the order of 10$^{-9}$ seconds.

In a typical ferromagnet, the magnitude of fluctuating fields can adopt a wide range of values.  For 
example, the field from spins on the manganese may add constructively or destructively at the lanthanum 
site, leading to a random distribution of fields among different lanthanum sites.  In the case where the size of the fluctuating field may adopt a range of values, the electron spin-spin correlation function should adopt the form of a gaussian.  An exponential spin-spin correlation function results if the fluctuating field can adopt one of only two values.  In the case of a gaussian correlation function, the spin-lattice relaxation time adopts the form:
\begin{equation}
\ln(\tau)\propto a+bH^2_0
\end{equation}	
which does not fit any of the data, except possibly at the highest temperature, where the uncertainties in the data are large.  The field dependence of nuclear spin-lattice relaxation thus reveals an unusual functional form for the electron spin-spin correlation function as well as unusually slow dynamics.  The explanation for this unusual behavior remains unclear.
\section{conclusions}

We have established the properties of $^{139}$La NMR of La$_{2/3}$Ca$_{1/3}$MnO$_{3}$ in the 
paramagnetic state which provide a basis for interpretation of $^{139}$La NMR measurements in the 
manganites. The spectrum is a standard powder pattern broadened to a significant degree by static lattice distortions and to a lesser degree by a distribution of susceptibility among lanthanum sites.  The static lattice distortions may be due to defects or inherent disorder due to Jahn-Teller distortions or calcium substitution.  In the temperature range studied, static lattice distortions around the lanthanum site do not change.  Also, changes in the hopping rate of localized field inhomogeneities, such as small polarons, do not affect the spectrum.  Measurements of nuclear spin-lattice relaxation at different points across the spectrum indicate that the relaxation mechanism is magnetic, implying that spin-lattice relaxation probes electron spin dynamics rather than lattice dynamics.  The field dependence of the spin-lattice relaxation reveals several unusual features about the magnetic fluctuations seen at the lanthanum site.  First, the fluctuations are extremely slow at all temperatures examined here.  Second, the field dependence of the fluctuations crosses over from T$_1\propto H_0^2$ near the Curie temperature to T$_1\propto H_0$ at 375K.  The physical origin of the unusual temperature and field dependence of the spin-lattice relaxation remains unclear.

\acknowledgments
	This work has been supported by the U.S. Department of Energy, Division of 
Materials Research under Grant DEFG02-91ER45439 through the University of Illinois at Urbana-Champaign, 
Frederick Seitz Materials Research Laboratory. C.P.S., M.J., and M.B.S. also acknowledge support from the National Science Foundation Grant No. DMR91-20000 through the Science and Technology Center for Superconductivity.
 	K.E.S thanks Nick Curro, Boris Fine, Jurgen Haase, Bob Heffner, Takashi Imai, Craig Milling, Dirk Morr, Jorg Schmalian, Dylan Smith, Raivo Stern, and Kazuyoshi Yoshimura for valuable assistance and comments.
\appendix
\section*{distinguishing magnetic and electric field gradient mechanisms of nuclear relaxation}

In an applied magnetic field, $H_0$, a spin-$7/2$ nucleus with magnetic quantum number $m=-7/2,-5/2,\ldots,5/2,7/2$ adopts energy levels with energies:
\begin{equation}
E_m=m\gamma_nH_0
\end{equation}
In thermal equilibrium, the population, $N_m$, of nuclei with energy $E_m$ obeys the Boltzmann distribution:
\begin{equation}
N_m\propto\exp\left(-\frac{E_m}{k_BT}\right)
\end{equation}
For the measurements described here, perturbations to the energies, $E_m$, due to magnetic or quadrupolar shifts are too small to have a significant effect on the populations of the energy levels.  The quadrupolar shifts are, however, large enough allow us to distinguish the position of the central transition from the satellites in our spectrum measurements.  Our measurements actually determine the population difference between adjacent levels.  For example, the size of the signal of the central transition (m=+1/2 to -1/2 transtion) is proportional to $N_{-1/2}-N_{1/2}$.
	
When measuring the spin-lattice relaxation time, $\tau$, one disturbs the populations of the energy levels and measures the rate at which the populations return to equilibrium.  In the so-called inversion-recovery measurements described in this paper, we use radiofrequency pulses to exchange the populations of two adjacent energy levels and then measure the populations of these levels during their return to equilibrium.  Because the broadening of the spectrum due to the quadrupole interaction is much greater than the bandwidth of the pulses, the chances of affecting more than one energy level of the same nucleus are small, even when effects due to magnetic broadening or a non zero asymmetry parameter are taken into account.
	
In the case in which relaxation occurs due to magnetic fluctuations, the populations of the energy levels recover from a disturbance according to a set of rate equations \cite{ANDREW,MARTINDALE}:
\begin{equation}
\frac{dN_m}{dt}=-W_mN_m + W_{m+1}N_{m+1} + W_{m-1}N_{m-1},
\end{equation}	
Where $W_m$ is the sum of matrix elements of the Hamiltonian describing the coupling between the magnetic fluctuations and the nucleus.  Because magnetic fluctuations can only induce transitions with $\Delta m=\pm1$, $W_m$ only includes matrix elements describing the coupling between levels m and m$\pm$1.  Defining the population difference between levels m and m+1:
\begin{equation}
n_{m,m+1}\equiv N_{m+1}-N_m,
\end{equation}
and solving the rate equations for the central transition in an inversion-recovery experiment, one finds:
\begin{eqnarray}
n_{-1/2,1/2}(t)&\propto& 1-\frac{1225}{858}\exp\left(-\frac{56t}{\tau}\right)-\frac{75}{182}\exp\left(-\frac{30t}{\tau}\right)\nonumber\\
& &-\frac{3}{22}\exp\left(-\frac{12t}{\tau}\right)-\frac{1}{42}\exp\left(-\frac{2t}{\tau}\right)
\end{eqnarray}
with different coefficients and time constants for each of the satellites.  It turns out that the signal, $n_{m,m+1}$(t), is well-approximated by single-exponential recovery for each of the transitions:
\begin{equation}
n_{m,m+1}(t)\propto 1-2\exp\left(-\frac{t}{T^m_1}\right).
\end{equation}
For the inversion-recovery experiment, these effective values relaxation time, $T^m_1$, adopt the following values:
\begin{center}
\begin{tabular}{cc}
\underline{  m  } & \underline{$T^m_1$}\\[0.5ex]
1/2 & 1.00 \\
3/2 & 1.10 \\
5/2 & 1.39 \\
7/2 & 2.58 
\end{tabular}
\end{center}
where we have set $T^{1/2}_1$ to 1.00 arbitary units.  The time constant $T^m_1$ corresponding to the central transition is shorter than those of the satellites, and $T^m_1=T^{-m}_1$.  To take into account the overlap of satellites, we average over the contributions of each satellite at any particular frequency, giving a relaxation time,$T^{average}_1(\nu)$ which depends of the frequency position on the spectrum:
\begin{eqnarray}
\lefteqn{1-2\exp\left(-\frac{t}{T^{average}(\nu)}\right)}\nonumber\\
&&=1-2\sum_mc_m(\nu)\exp\left(-frac{t}{T^m_1}\right).
\end{eqnarray}
where the constants $c_m(\nu)$ give the intensity of the transition between levels m and m+1 at the frequency $\nu$, a quantity extracted from the fit to the spectrum.

In the case in which relaxation occurs due to electric field gradient fluctuations, both $\Delta m=\pm 1$ and $\Delta m=\pm 2$ transtions are allowed, giving rate equations of the form:
\begin{eqnarray}
\frac{dN_m}{dt}&=&-W_mN_m + W_{m+1}N_{m+1} + W_{m-1}N_{m-1}\nonumber\\
& &+W_{m+2}N_{m+2}+W_{m-2}N_{m-2},
\end{eqnarray}
In a typical system, the matrix elements with $\Delta m=\pm 1$ and $\Delta m=\pm 2$ are nearly the same \cite{ANDREW}.  In this case, the relaxation again adopts a nearly single-exponential form.  In contrast with the case of magnetic relaxation, the relaxation time of the central transition is longer than those of the satellites.  It is possible for the opposite trend to occur in the situation in which electric field gradient fluctuations dominate, but only if the $\Delta m=\pm 2$ matrix elements are over an order of magnitude larger than the $\Delta m=\pm 1$ matrix elements.

\clearpage
\begin{figure}
\caption{Comparison of 575K $^{139}$La NMR spectra measured at 8.19 (filled squares) and 4.00 Tesla 
(open circles).  Lines are guides to the eye.  The frequency axes have been shifted for the sake of 
comparison.  Inset:  Comparison of narrow part of the spectrum.  The frequency scale has been multiplied 
by the applied field to show that the width of the narrow part scales inversely with applied field.}
\label{HTSPECTRUM}
\end{figure}

\begin{figure}
\caption{Comparison of calculated powder pattern with no adjustable parameters and the 575K, 8.19 T spectrum.  Inset on left is a magnified view of the narrow central feature.  Inset on the right is a magnified view of the right side of the broad feature.}
\label{UNBROADFIT}
\end{figure}

\begin{figure}
\caption{Same as figure~\ref{UNBROADFIT}, except the calculated powder pattern has been broadened by a distribution of magnetic shifts}
\label{MAGBROADFIT}
\end{figure}

\begin{figure}
\caption{Same as figure~\ref{UNBROADFIT}, except the calculated powder pattern has been broadened by a distribution of electric field gradients and magnetic shifts.}
\label{EFGBROADFIT}
\end{figure}

\begin{figure}
\caption{Spectra measured at several temperatures.  The areas have been normalized to account for $1/T$ 
Curie law behavior of the nuclear polarization and for changes in the Q of  the detection circuit.}
\label{TEMPDEP}
\end{figure}

\begin{figure}
\caption{Plot of (resonant frequency)/(shift) versus temperature.  The line is a fit to Curie-Weiss behavior with a Curie temperature of 269$\pm$19K.}
\label{CURIE}
\end{figure}

\begin{figure}
\caption{Room temperature spectrum (filled squares) and broad part of 575K spectrum (open circles) 
measured at 8.19 Tesla.  Frequencies have been shifted and amplitudes scaled for the sake of comparison.}
\label{EFGINDEPT}
\end{figure}

\begin{figure}
\caption{Temperature dependence the of inverse of the width of the convolution gaussian used to fit the 
spectra.  The line is a fit to Curie-Weiss behavior with a Curie temperature of 280$\pm$30K.}
\label{CURIEWIDTH}
\end{figure}

\begin{figure}
\caption{Fast Fourier Transforms of spin echoes measured a time, $T_{ev}$ after selectively inverting portions of the spectrum.  Measurements shown were performed at room temperature in a 4 Tesla field.}
\label{SWAVE}
\end{figure}

\begin{figure}
\caption{Spin-lattice relaxation times measured at 575K at 9.40 Tesla at different positions across the 
spectrum.  Zero shift corresponds to the apex of the central transition.  Times have been normalized to the value measured at zero shift.  The line is a fit which assumes magnetic relaxation and overlap among 
transitions as determined from the fit in figure~\ref{EFGBROADFIT}.}
\label{T1VSPECT}
\end{figure}

\begin{figure}
\caption{Spin lattice relaxation time at room temperature versus $H_0^2$.  The fit corresponds to behavior expected for a system with an exponential electron spin-spin correlation function as described in the text.}
\label{T1VH}
\end{figure}

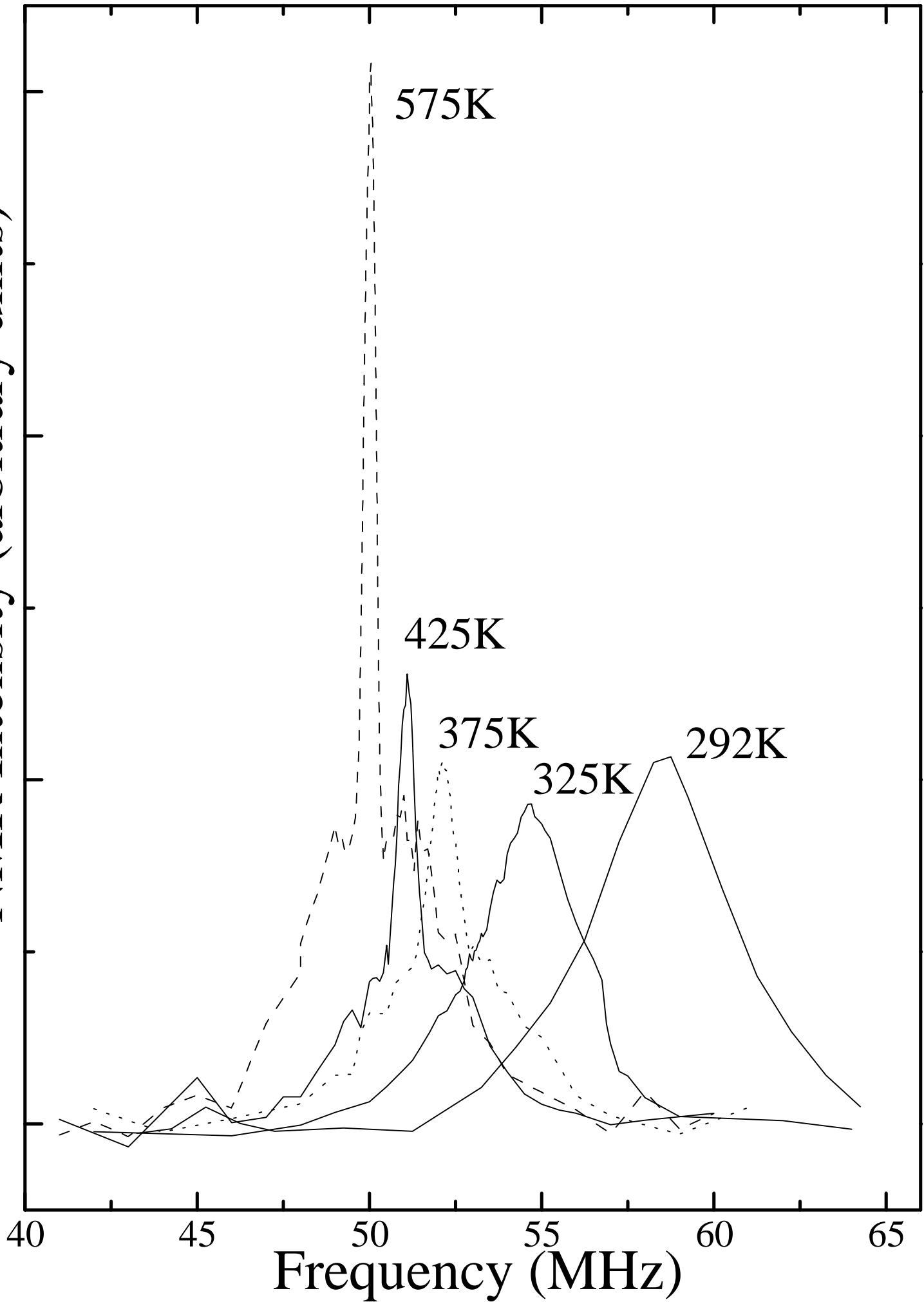
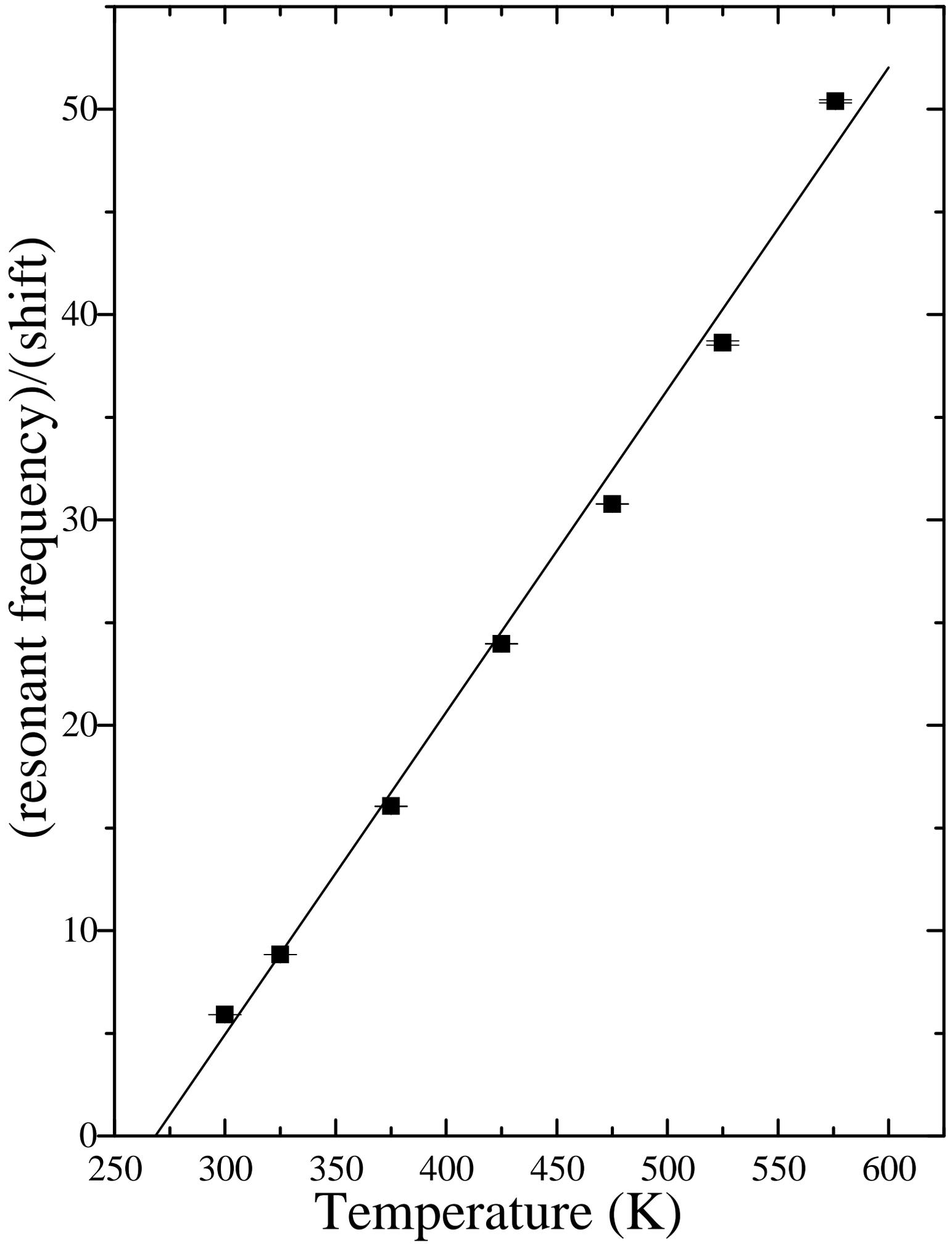
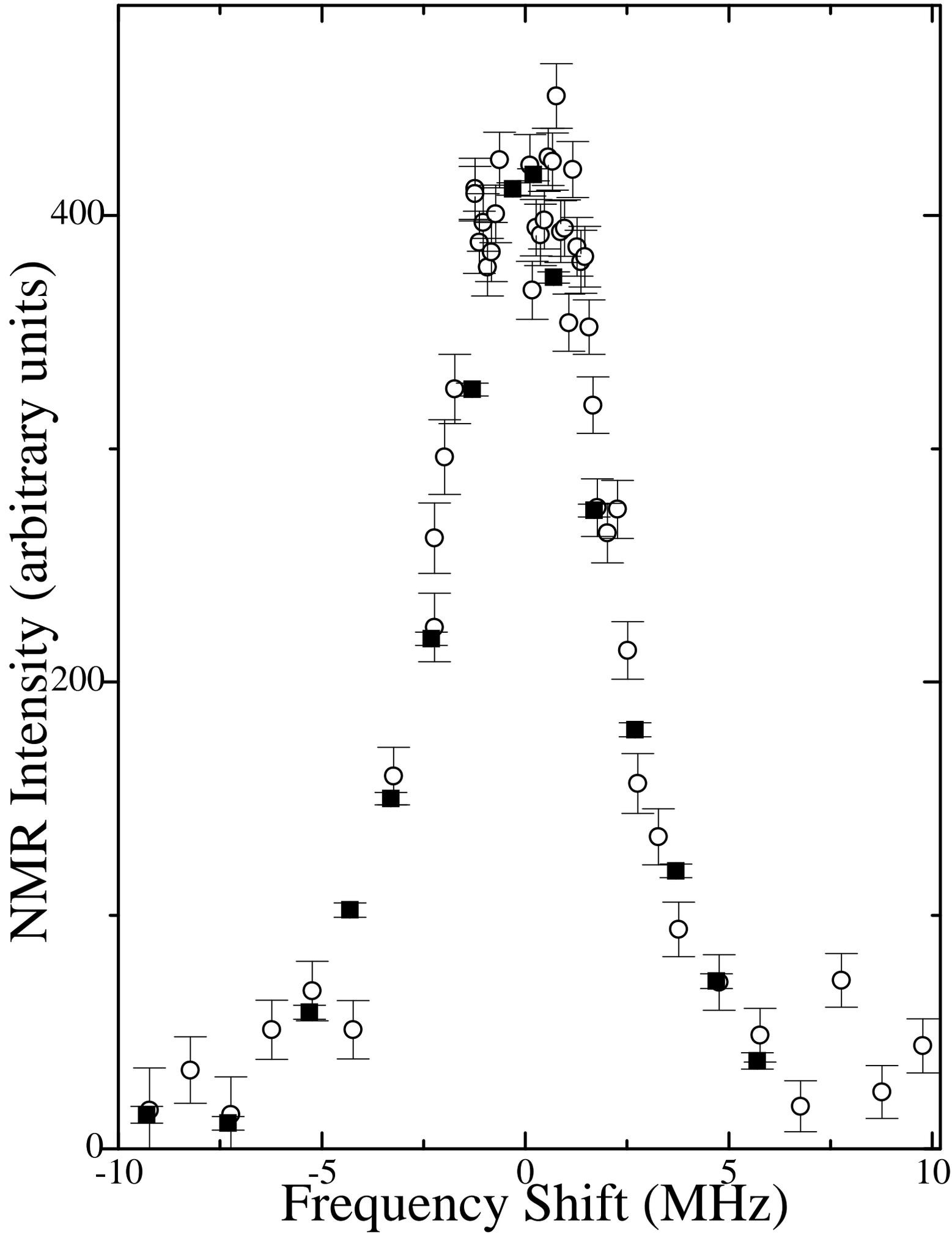
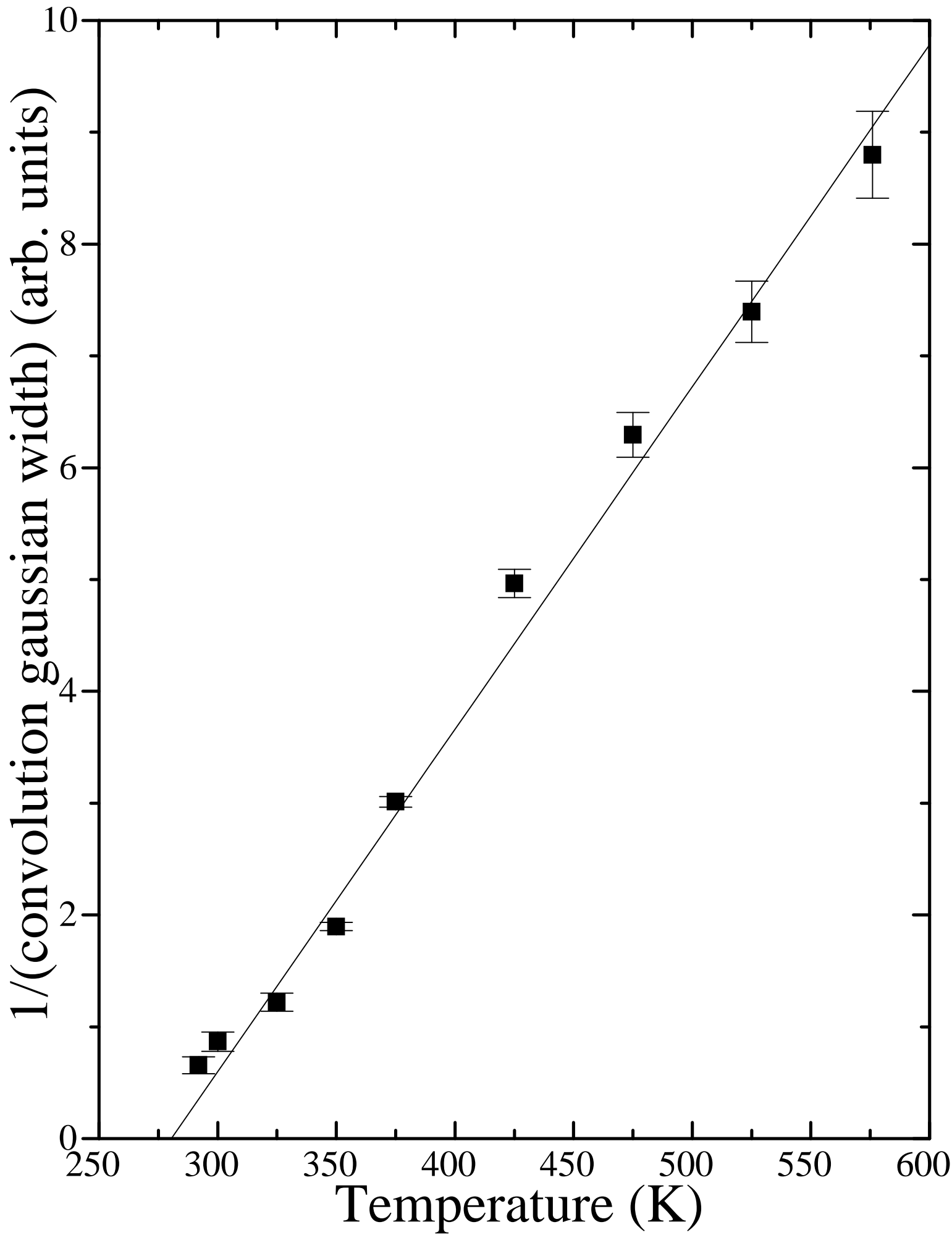
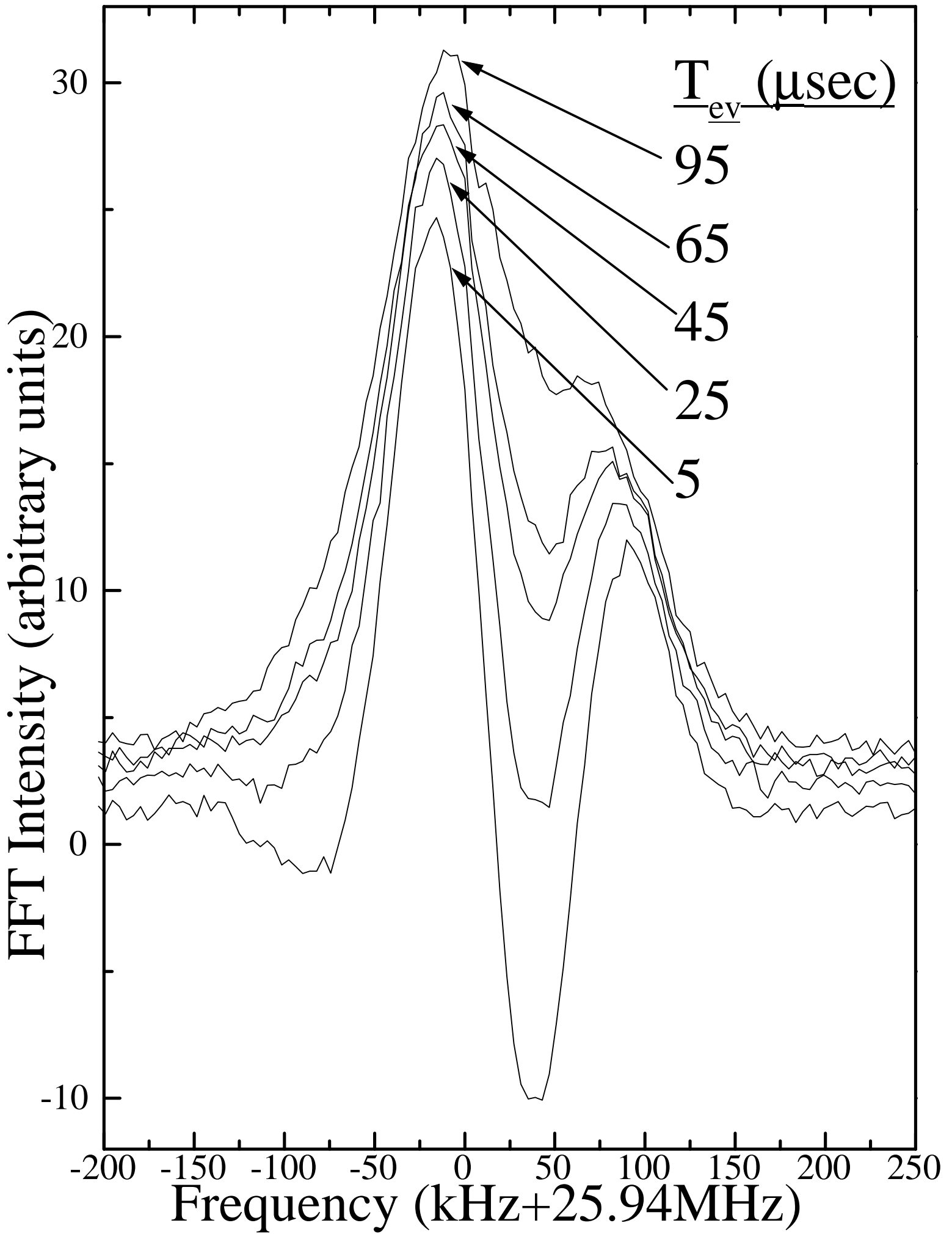
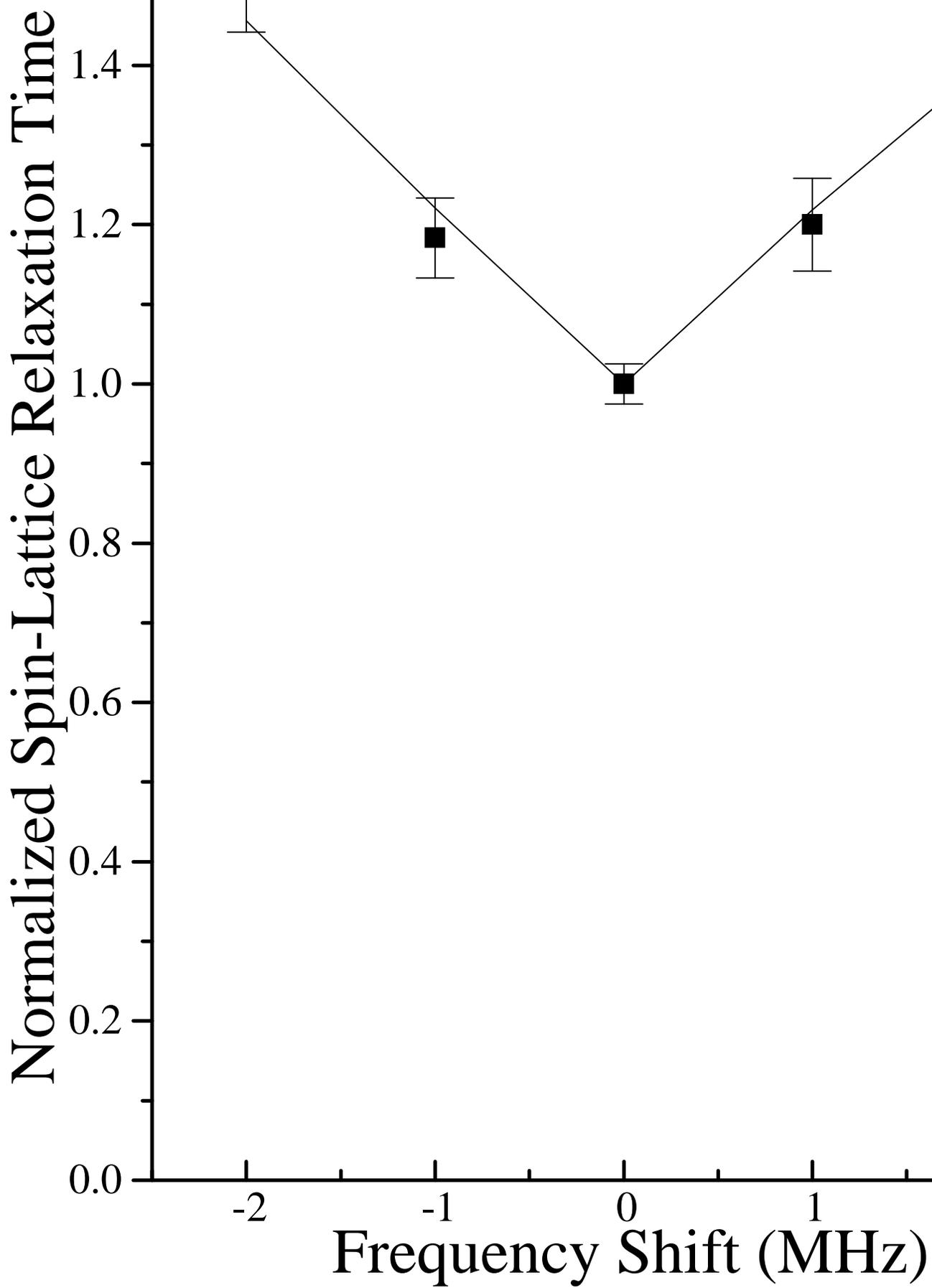
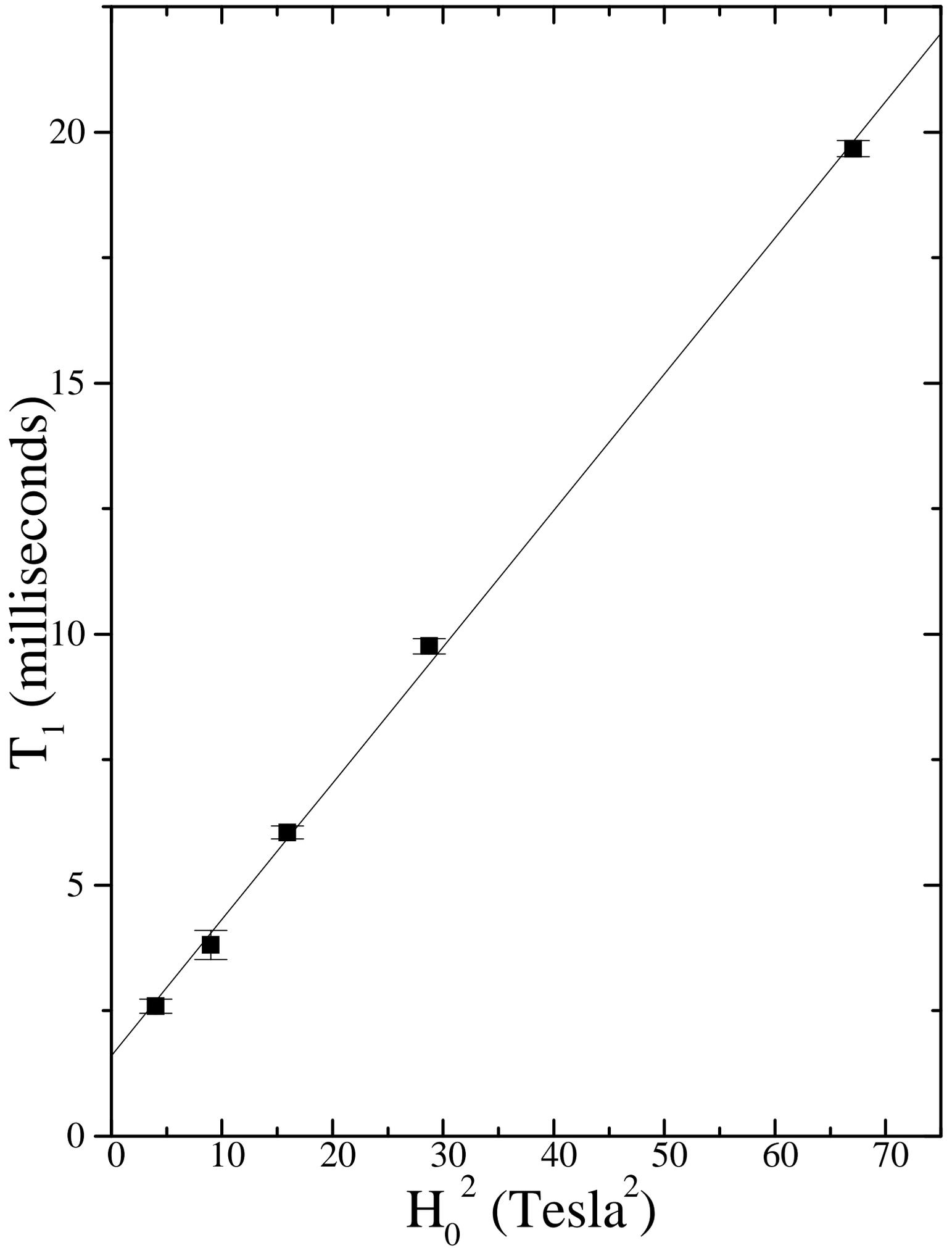
\end{document}